\documentstyle[12pt]{article}
\begin{document}
\bigskip
\bigskip
\bigskip
\bigskip
\bigskip
\bigskip
\bigskip
\begin{center}
{\large \bf Current Algebra: Quarks and What Else?}
\end{center}
\bigskip
\begin{center}

{\bf Harald Fritzsch\footnote[1]{On leave from the Max--Planck--Institut
f\"ur Physik und Astrophysik. M\"unchen, Germany.}\footnote[2]{Present
address: Lauritsen Laboratory of High Energy Physics, California,
Institut of Technology, Pasadena, California.}}
\end{center}
\begin{center}
{\bf and}
\end{center}
\begin{center}
{\bf Murray Gell--Mann\footnote[7]{John Simon Guggenheim Memorial
Foundation Fellow.}$^\dagger$}
\end{center}
\begin{center}
{\bf CERN, Geneva, Switzerland}\\
\end{center}
\bigskip
\begin{center}
Proceedings of the XVI International Conference on High Energy
Physics, Chicago, 1972. Volume 2, p. 135 (J. D. Jackson, A. Roberts, eds.)
\end{center}

\bigskip
\bigskip
\bigskip
\bigskip
\bigskip
\begin{abstract}
  After receiving many requests for reprints of this article,
  describing the original ideas on the quark gluon gauge theory, which
  we later named QCD, we decided to place the article in the e--Print
  archive.
\end{abstract}
\newpage
\noindent
{\bf I. Introduction}\\
{\small For more than a decade, we particle theorists have been squeezing
predictions out of a mathe\-matical field theory model of the hadrons that
we don't fully believe -- a model containing a triple of spin 1/2 fields
coupled universally to a neutral spin 1 field, that of the ``gluon''. In
recent years, the triplet is usually taken to be the quark triplet, and it
is supposed that there is a transformation, presumably unitary, that
effectively converts the current quarks of the relativi\-stic model into the
constituent quarks of the naive quark model of baryon and meson spectrum
and couplings.\\
We abstract results that are true in the model to all orders of the
gluon coupling and postulate that they are really true of the
electromagnetic and weak currents of hadrons to all orders of the strong
interaction. In this way we build up a system of algebraic relations,
so--called current algebra, and this algebraic system gets larger and larger
as we abstract more and more properties of the model.\\
In section III, we review briefly the various stages in the history of
current algebra. The older abstractions are correct to each order of
renormalized perturbation theory in the model$^{1)}$, while the more recent ones,
those of light cone current algebra, are true to all orders only
formally$^{3)}$.
We describe the results of current algebra$^{2)}$ in terms of commutators on or
near a null plane, say $x_3 + x_0 = 0$.\\
In section IV, we attempt to describe, in a little more detail, using null
plane language, the system of commutation relations valid in a free quark
model that are known to remain unchanged (at least formally) when the
coupling to a vector ``gluon'' is turned on. These equations give us a
formidable body of information about the hadrons and their currents,
supposedly exact as far as the strong interaction is concerned, for
comparison with experiment. However, they by no means exhaust the degrees
of freedom present in the model; they do not yield an algebraic
system large enough to contain a complete description of the hadrons.
In an Appendix, the equations of Section IV are related to form factor
algebra.\\
In Section V, we discuss how further commutation of the physical
quantities arising from light cone algebra leads, in the model field
theory, to results dependent on the coupling constant, to formulae in which
gluon field strength operators occur in bilocal current operators
proliferate. Only when these relations are included do we finally get an
algebraic system that contains nearly all the degrees of freedom of the model.
We may well ask, however, whether it is the right algebraic system.
We discuss briefly how the complete description of the hadrons involves
the specification and slight enlargement of this algebraic system, the
choice of representation of the algebra that corresponds to the complete
set of hadron states, and the form of the mass or the energy operator,
which must be expressible in terms of the algebra when it is complete.
The choice of representation may be dictated by the algebra, and
if so that would justify the use of a quark and gluon Fock space by
some ``parton'' theorists.\\
Finally, in Section VI, it is suggested that perhaps there are alternatives
to the vector gluon model as sources of information or as clues for the
construction of the true hadron theory. Assuming we have described the
quark part of the model correctly, can we replace the gluons by something
else? The ``string'' or ``rubber band'' formulation, in ordinary
coordinate space, of the zeroth approximation to the dual resonance model,
is suggested as an interesting example.\\
\\
Before embarking on our discussion of current algebra, we discuss in
Section II the crucial point that quarks are probably not real particles
and probably obey special statistics, along with related matters
concerning the gluons of the field theory model.}\\
\\

\noindent
\underline{II. FICTITIOUS QUARKS AND ``GLUONS'' AND THEIR STATISTICS}\\
\\
We assume here that quarks do not have real counterparts that are
detectable in isolation in the laboratory -- they are supposed to be
permanently bound inside the mesons and baryons. In particular, we assume
that they obey the special quark statistics, equi\-valent to ``para--Fermi
statistics of rank three'' plus the requirement that mesons always be
bosons and baryons fermions. The simplest description of quark statistics
involves starting with three triplets of quarks, called red, white, and
blue, distinguished only by the parameter referred to as color. These nine
mathematical entities all obey Fermi--Dirac statistics, but real particles
are required to be singlets with respect to the $SU_3$ of color, that is
to say combinations acting like
\begin{equation}
\bar{q}_R q_R + \bar{q}_B q_B + \bar{q}_W q_W \, \, \, {\rm or} \, \, \,
q_R q_B q_W - q_B q_R q_W - q_R q_W q_B - q_W q_B q_R + q_W q_R q_B +
q_Bq_Wq_R \, .
\end{equation}
The assumption of quark statistics has been common for many years, although
not ne\-cessarily described in quite this way, and it has always had the
following advantage: The constituent quarks as well as current quarks would
obey quark statistics, since the transformation between them would not
affect statistics, and the constituent quark model would then assign the
lowest-lying baryon states (56 representation) to a symmetrical spatial
configuration, as befits a very simple model.\\
Nowadays there is a further advantage. Using the algebraic relations
abstracted formally from the quark--gluon model, one obtains a formula for
the $\pi^0$ decay amplitude in the PCAC approximation, one that works
beautifully for quark statistics but would fail by a factor 3 for a single
Fermi--Dirac triplet$^{4)}$.\\
We have the option, no matter how far we go in abstracting results from
a field theory model, of treating only color singlet operators. All the
currents, as well as the stress--energy--momentum tensor
$\Theta_{\mu \nu}$
that couples to gravity and defines the theory, are color singlets. We
may, if we like, go further and abstract operators with three quark fields,
or four quark fields and an antiquark field, and so forth, in order to
connect the vacuum with baryon states, but we still need select only
those that are color singlets in order to connect all physical hadron
states with one another.\\
It might be a \underline{convenience} to abstract quark operators
themselves, or other non--singlets with respect to color, along with
fictitious sectors of Hilbert space with triality non--zero, but it is
not a \underline{necessity}. It may not even be much of a convenience,
since we would then, in describing the spatial and temporal variation
of these fields, be discussing a fictitious spectrum for each fictitious
sector of Hilbert space, and we probably don't want to load ourselves with
so much spurious information.\\
We might eventually abstract from the quark--vector--gluon field theory
model enough algebraic information about the color singlet operators in
the model to describe all the degrees of freedom that are present.\\
For the real world of baryons and mesons, there must be a similar
algebraic system, which may differ in some respects from that of the model,
but which is in principle knowable. The operator $\Theta_{\mu \nu}$ could
then be expressed in terms of this system, and the complete Hilbert space
of baryons and mesons would be a representation of it. We would have a
complete theory of the hadrons and their currents, and we need never
mention any operators other than color singlets.\\
Now the interesting question has been raised lately whether we should
regard the gluons as well as the quarks as being non--singlets with
respect to color$^{5)}$. For example, they could form a color octet of neutral
vector fields obeying the Yang--Mills equations. (We must, of course,
consider whether it is practical to add a common mass term for the gluon
in that case -- such a mass term would show up physically as a term in
$\Theta_{\mu \nu}$ other than the quark bare mass term. In the past, we
have referred to such an additional term that violates scale invariance,
but does not violate $SU_3 \times SU_3$ as $\delta $ and its dimension
as $l_{\delta }$. Nowadays, ways of detecting expected values of $\delta $
are emerging.)$^{6)}$.\\
If the gluons of the model are to be turned into color octets, then an
annoying asymmetry between quarks and gluons is removed, namely that there
is no physical channel with quark quantum numbers, while gluons communicate
freely with the channel containing the $\omega $ and $\phi $ mesons. (In
fact, this communication of an elementary gluon potential with the real
current of baryon number makes it very difficult to believe that all the
formal relations of light cone current algebra could be true even in a
``finite'' version of singlet neutral vector gluon field theory.)\\
If the gluons become a color octet, then we do not have to deal with a
gluon field strength standing alone, only with its square, summed over the
octet, and with quantities like $\bar{q} \left( \partial_{\mu} - igo _A
B_{A\mu} \right) q$, where the $\sigma $'s are the eight $3 \times 3$
color matrices for the quark and the B's are the eight gluon potentials.\\
\\
Now, suppose we look at such a model field theory, with colored quarks and
colored gluons, including the stress--energy--momentum tensor. Basically the
questions we are asking are the following:
\begin{enumerate}
\item[1.] Up to what point does the algebraic system of the color singlet
operators for the real hadrons resemble that in the model? What is it in
fact?
\item[2.] Up to what point does the representation of the algebraic system
by the Hilbert space of physical hadron states resemble that in the
model? What is it in fact?
\item[3.] Up to what point does $\Theta_{\mu \nu}$, expressed in term of
the algebraic system, resemble that in the model? What is it in fact?
\end{enumerate}
\bigskip
The measure of our ignorance is that for all we know, the algebra of
color singlet operators, the representation, and even the form of
$\Theta_{\mu \nu}$ could be exactly as in the model! We don't yet know
how to extract enough consequences of the model to have a decisive
confrontation with experiment, nor can we solve the formal equations for
large $g$.\\
If we were solving the equations of a model, the first question we would
ask is: Are the quarks really kept inside or do they escape to
infinity? By restricting physical states and interesting operators to color
singlets only, we have to some extent begged that question. But it
re-emerges in the following form:\\
With a given algebraic system for the color singlet operators, can we
find a locally causal $\Theta _{\mu \nu}$ that yields a spectrum
corresponding to mesons and baryons and antibaryons and combinations
thereof, or do we find a spectrum (in the color singlet states) that
looks like combinations of free quarks and antiquarks and gluons?\\
In the next three Sections we shall usually treat the vector gluon,
for convenience, as a color singlet.\\
\\
\\
\underline{III. REVIEW OF CURRENT ALGEBRA}\\
\\
In this section we sketch the gradual extension of algebraic results
abstracted from free quark theory that remain true, either in
renormalized perturbation theory or else only formally, when the coupling
to a neutral vector gluon field is turned on.\\
The earlier abstractions were of equal--time commutation relations of
current components. It was soon found that useful sum rules could best be
derived from these by taking matrix elements between hadron states of
equal $P_3$ as $P_3 \rightarrow \, \infty$, selecting the ``gluon''
components of
the currents (those with matrix elements finite in this limit rather than
tending to zero), and adding the postulate that, in the sum over
intermediate states in the commutator, only states of finite mass need be
considered. Thus formulae like the Adler--Weisberger and Cabibbo--Radicati
sum rules were obtained and roughly verified by experiment.\\
Nowadays, the same procedure is usually accomplished in a slightly different
way that is a bit cleaner -- the hadron momenta are left finite instead
of being boosted by a limit of Lorentz transformations, and the equal time
surface is transformed by a corresponding limit of Lorentz transformations
into a null plane, with $x_3 + x_0 = $ constant, say zero. The hypothesis
of saturation by finite mass intermediate states is replaced by the
hypothesis that the commutation rules of good components can be abstracted
from the model not only on an equal time plane, but on a null plane as
well$^{7, 8)}$.\\
In the last few years, the process of abstraction has been extended to a
large class of algebraic relations (those of ``light cone current
algebra'') that are true only formally in the model, but fail to each
order of renormalized perturbation theory - they would be true to each
order if the model were super--renormalizable. The motivation has been
supplied by the compatibility of the deep inelastic electron scattering
experiments performed at SLAC with the scaling predictions of Bjorken,
which is the most basic feature of ``light cone current algebra''. The
Bjorken scaling limit $q^2 \rightarrow \infty, 2p \cdot q \rightarrow
\infty, \xi  \equiv q^2/\left( -2p \cdot q \right)$ finite) corresponds in
coordinate
space to the singularity on the light cone $\left( x-y \right)^2 = 0$ of
the current commutator $\left[ j(x), j(y) \right]$, and the relations of
light cone current algebra are obtained by abstracting the leading
singularity on the light cone from the field theory model. The singular
function of $x-y$ is multiplied by a bilocal current operator
$\Theta \left( x, y \right)$ that reduces to a familiar local
current as $x - y \rightarrow 0$. The Bjorken scaling functions
$F \left( \xi \right)$ are Fourier transforms of the expected values of
the bilocal operators. Numerous predictions emerge from the relations
abstracted from the quark--gluon model for deep inelastic and neutrino
cross--sections. For example, the spin 1/2 character for the quanta bearing
the charge in the model is reflected in the prediction
$\sigma_L / \sigma_T \rightarrow 0$, while the charges of the
quarks are reflected in the inequalities
$1/4 s F^{\rm en} \left( \xi \right) /F^{\rm ep} (\xi) \le 4$. So far
there is no clear sign of my contradiction between the formulae and the
experimental results.\\
We may go further and abstract from the model also the light--cone
commutators of bilocal currents, in the limit in which all the intervals
among the four points approach zero, that is to say, when all four points
tend to lie on a light--like line. The same bilocal operators then recur as
coefficients of the singularity, and the algebraic system closes.\\
The light cone results can be reformulated in terms of the null plane.
We consider a commutator of local currents at two points $x$ and $y$
and allow the two points to approach the same null plane, say
\begin{equation}
x_+ \equiv x_3 + x_0 = 0, y_+ \equiv y_3 + y_0 = 0
\end{equation}
\\
As mentioned above, when both current components are ``good'', we obtain
a local commutation relation on the null plane, yielding another good
component, or else zero. But when neither component is good, there is a
singularity of the form
\begin{equation}
\delta \left( x_+ - y_+ \right)
\end{equation}
and the coefficient is a bilocal current on the null plane. It is this
singularity, arising from the light--cone singularity, that gives the
Bjorken scaling.\\
On the null plane, with $x_+ = 0$, the three coordinates are the transverse
spacelike coordinates $x_1$ and $x_2$ (called $x_{\perp}$) and the
lightlike coordinate $x_-  \equiv x_3 - x_0 $. Our bilocal currents
$O \left( u, y \right) $ on the nullplane are functions of four
coordinates: $x_-, y_-$ and $x_{\perp} = y_{\perp}$, since the interval
between $x$ and $y$ is lightlike.\\
We may now consider the commutator of two bilocal currents defined on
neighboring null planes (in each case with a lightlike interval between
the two arguments of the bilocal current). Again, when neither current
component is good, there is a $\delta $--function singularity of the
spacing between the two null planes and the coefficient is a bilocal
current defined on the common limiting null plane. In this language, as
before in the light cone language, the system of bilocal currents
closes.\\
We may commute two good components of bilocal currents on the same null
plane, and, as for local currents, we obtain a good component on the
right--hand side, without any $\delta $--function singularity at
coincidence of the two null planes. Thus the good components of the
bilocal currents $O (u, y)$ form a Lie algebra on the null plane,
a generalization of the old Lie algebra of local good components on the
null plane (recovered by putting $x_- = y_-$).\\
Now, how far can we generalize this new Lie algebra on the null plane
and still obtain exact formulae, formally true to all orders in the
coupling constant, but independent of it, so that free quark formulae
apply?\\
In the next section, we take up that question, but first we summarize the
situation of current algebra on and near the null plane.\\
\\
\\
\underline{IV. SUMMARY OF LIGHT CONE AND NULL PLANE RESULTS}\\
\\
Let us now be a little more explicit. We are dealing with 144 bilocal
quantities ${\cal F}_{j \alpha}, {\cal F}_{j \alpha}, S_j, P_j$ and
$T_{j \alpha \beta}$ all functions of $x-y$ with $\left( x - y
\right)^2 \rightarrow 0$. Let us select the 3--direction for our null
planes. Then in the model we can set $B_+ \equiv B_3 + B_0 = 0$ for the
gluon potential by a choice of gauge. The gauge--invariance factor exp
ig $\int\limits_{y}^{x} B \cdot dl$ for a straight line path on a null
plane is just exp
$\left[ i \frac{g}{2} B_+ \left( x_- - y_- \right) \right] = 1$.
Thus we have simple correspondences between our quantities
and operators in the model:
\begin{displaymath}
{\cal F}_{j \alpha} (x, y) \sim \frac{i}{2} \bar{q} (x) \lambda_j
\gamma_{\alpha} q (y), \, \, \, {\rm etc.}
\end{displaymath}

and we have introduced the notation ${\cal D} \left( x, y, \frac{i}{2}
\, \lambda_j \, \gamma_{\alpha} \right)$, etc., where
\begin{equation}
{\cal D} (x, y, G) \sim \bar{q} (x) Gq (y) \sim q^+ (x) (\beta G) q (y) \, .
\end{equation}
\\
We are dealing with ${\cal D} (x, y, G)$ for every $(12 \times 12)$ matrix
$G$, with
\begin{eqnarray}
{\cal F}^5_{j \alpha} (x, y) & = & {\cal D} \left( x, y, \frac{i}{2}
\lambda_j \gamma_{\alpha}, \gamma_5  \right) S_j (x, y) =
{\cal D} \left( x, y, \frac{1}{2} \, \lambda_j \right),  \\ \nonumber \\
P_j (x, y) & = & {\cal D} \left( x, y, \frac{i}{2} \lambda _j \,
\gamma_5 \right), \, \, \, {\rm and} \, \, \,
{\cal T}_{j \alpha \beta} (x, y) =
{\cal D} \left( x, y, \frac{i}{2} \, \lambda_j \, \sigma_{\alpha \beta}
\right).
\end{eqnarray}
\\
The good components, in the old equal--time $P_3 \rightarrow \infty$
language, were those with finite matrix elements between states of finite
mass and $P_3 \rightarrow \infty$. By contrast, bad components were those
with matrix elements going like $P_3 \, ^{-1}$ and terrible components
those with matrix elements going like $P_3 \, ^{-2}$.\\
\\
In the null plane language, good components are those for which $\beta G$
is proportional to $1 + \alpha_3$; thus the 36 good components are
${\cal F}_{j+}, {\cal F}_{j+}^5, {\cal T}_{j1+}, {\cal T}_{j2+}$ for
$j = 0 \ldots 8$. The terrible components are those for which $\beta G$
is proportional to $1 - \alpha_3$, hence ${\cal F}_{j-}, {\cal F}_{j-}^5,
{\cal T}_{j1}$, and ${\cal T}_{j2-}$. The rest are bad; they have
$\beta G$ anticommuting with $\alpha_3$ so that $\alpha _3$ is -1 on the
left and $+1$ on the right or vice versa.\\
Now the leading light cone singularity in the commutator of two bilocals
is just given by the formula
\begin{equation}
\left[ \left( {\cal D} \left( x, y, G \right), {\cal D} \left( u, v, G'
\right) \right) \right] \hat{=} {\cal D} \left( x, v, iG \gamma_{\mu}
G' \right) \partial_{\mu} \Delta (y - u) - {\cal D}
\left( u, y, iG' \gamma_{\mu} G \right) \partial_{\mu} \Delta (v-x),
\end{equation}
\\
with $\Delta (z) = (2 \pi )^{-1} \, \varepsilon \left( z_0 \right)
\delta \left( z^2 \right)$.\\
\\
When we commute two operators with coordinates lying on neighboring null
planes with separation $\Delta x_+$, a singularity of the type $\delta
\left( \Delta x_+ \right)$ appears (as we have mentioned in Section III)
multiplied by a bilocal operator, with coordinates lying in the common
null plane as $\Delta x_+ \rightarrow 0$, and it is this term that gives
rise to Bjorken scaling. The term in question comes from the component
$\frac{\partial}{\partial z_+} \Delta (z)$ in $\partial_{\mu} \Delta
(z)$, and is thus multiplied by ${\cal D} \left( x, \nu, i G \gamma_+
G' \right)$ and ${\cal D} \left( u, y, iG' \gamma_+ G \right)$. Now
$\beta \left( iG \gamma_+ G' \right) = \left( \beta G \right) \left(
1-\alpha_3 \right) \left(\beta G' \right)$, so it is clear that the
singular Bjorken scaling term vanishes for good--good and good--bad
commutators. In the case of the other components, we have, schematically,
[bad, bad] $\rightarrow $ good, [bad, terrible] $\rightarrow $ bad, and
[terrible, terrible] $\rightarrow $ terrible for the Bjorken singularity.\\
The vector and axial vector local currents ${\cal F}_{j \alpha} (x, x)$
and ${\cal F}_{j \alpha}^5 (x, x)$ occur, of course, in the electromagnetic
and weak interactions. The local scalar and pseudoscalar currents occur
in the divergences of the non--conserved vector and the axial vector
currents, with coefficients that are linear combinations of the bare quark
masses, $m_u, m_d$ and $m_s$, treated as a diagonal matrix. (Here $m_u$
would equal $m_d$ if isotopic spin conservation were perfect, while the
departure of $m_s$ from the common value of $m_u$ and $m_d$ is what gives
rise to $SU_3$ splitting; the non--vanishing of $m$ is what breaks
$SU_3 \times SU_3$).\\
We see that all the 144 bilocals are physically interesting, including
the tensor currents, because they all occur in the commutators of these
local $V, A, S$, and $P$ densities as coefficients of the $\delta
\left( \Delta x_+ \right)$ singularity. Commuting a local scalar with
itself or a local pseudoscalar with itself leads to the same bilocal as
commuting a transverse component of a vector with itself, and thus the
light cone commutator of current divergences is predicted to lead to
Bjorken scaling functions that are proportional to those observed in
the light cone commutation of currents, while the coefficients permit
the experimental determination of the squares of the quark bare masses.
Unfortunately, the relevant expe\-riments are difficult. (The finiteness
of the bare masses, as compared with the divergences encountered term in
renormalized perturbation theory in a gluon model, presumably has the same
origin as the scaling, which also fails term by term in renormalized
perturbation theory.)\\
As we have outlined in Section III, we begin the construction of the
algebraic system on the null plane by commuting the good bilocals with
one another. The leading singularity on the light cone (Eq.(4.1)) gives
rise to the simple closed algebra we have mentioned, but we need also the
additional assumption that lower singularities on the light cone give no
contribution to the good--good commutators on the null plane. This
additional assumption can be squeezed out of the model in various ways.
The simplest, however, is to use canonical quantization of the quark--gluon
model on the null plane.\\
In the model, the quark field $q$ is written as $q_+ \, + q_-$, where
$q_{\pm} = \frac{1}{2}$ $ \left( 1 \pm \alpha_3 \right) q$.
Then $q_+$ obeys the canonical rules
$\left\{ q_{+ \alpha} (x), q_{+\beta} (y) \right\} = 0,
\left\{ q_{+ \alpha} (x). q^+_{+ \beta} (y) \right\} = \delta^{(3)} (x-y)
\frac{1}{2} \left( l + a_3 \right)_{\alpha \beta}$ on the null plane, where
$\delta^{(3)} (x-y) = \delta \left( x_{\perp} - y_{\perp} \right) \delta
\left( x_- - y_- \right)$. Thus for any good matrices $\beta A_{++}$ and
$ \left( \beta B_{++}  \right) $, we have on the null plane

\begin{displaymath}
\left[ {\cal D} \left( x, y, \beta A_{++} \right), {\cal D} \left(
u, v, \beta B_{++} \right) \right] =
\end{displaymath}
\begin{displaymath}
{\cal D} \left( x, v \beta A_{++}
B_{++} \right) \delta^{(3)} (y-u) - {\cal D} \left( u, y,
\beta B_{++}A_{++} \right) \delta^{(3)} (v-x),
\end{displaymath}
\noindent
which is just what we would get from (4.1) with no additional contribution
from lower light cone singularities.\\
The good--good commutation relations (4.2) on the null plane, together
with the equations (4.1) for the leading light cone singularity in the
commutator of two bilocal currents, illustrate how far we can go with
abstracting free quark formulae that are formally unchanged in the model
when the gluon coupling is turned on.\\
One may go further in certain directions. For example, the formulae for
the leading light cone singularity presumably apply to disconnected as
well as connected parts of matrix elements, and thus the question of the
vacuum expected value of a bilocal operator arises. In the model, the
coefficient of the leading singularity as $\left( x-y \right)^2
\rightarrow 0$ of such an expected value is formally independent of the
coupling constant, and we abstract that as well -- the answer here is
dependent on statistics, however, and we assume the validity of quark
statistics. Thus we obtain predictions like the following:
\begin{equation}
\sigma \left( e^+ + e^- \rightarrow \, \, \, {\rm hadrons} \right) /
\sigma \left( e^+ + e^- \rightarrow \mu^+ + \mu^- \right) \rightarrow 2
\end{equation}
at high energy to lowest order in the fine structure constant.\\
\\
The leading light cone singularity of an operator product, or of a
physical order $\left( T^* \right)$ product, may also be abstracted from
the model, except for certain subtraction terms (often calculable and / or
unimportant) that behave like four--dimensional $\delta $--functions in
coordinate space. To go from a commutator formula to a physical ordered
product formula, we simply perform the substitutions
\begin{equation}
\left( 2 \pi \right)^{-1} \varepsilon (z) \delta \left( z^2 \right)
\rightarrow \left( 4 \pi^2 i \right)^{-1} \left( z^2 - i z_0 \varepsilon
\right)^{-1} \rightarrow \left( 4 \pi^2 i \right)^{-1}
\left( z^2 - i \varepsilon \right)^{-1} \, .
\end{equation}
\\
With the aid of the product formulae and the vacuum expected values, we
obtain the PCAC value of the $\pi^0 \rightarrow 2 \gamma $ decay
amplitude.\\
\\
Other exact abstractions from the vector gluon model that do not contain
$g$ are divergence and curl relations for local $V$ and $A$ currents:
\begin{eqnarray}
\frac{\partial}{\partial x_{\mu}} {\cal D} \left( x, x, \frac{i}{2}
\, \lambda_i \, \gamma_{\mu} \right) & = & {\cal D}
\left( x, x, \frac{i}{2}
\left[ m, \lambda_i \right] \right),
\nonumber \\ \nonumber \\
\frac{\partial}{\partial x_{\mu}} {\cal D} \left( x, x, \frac{i}{2} \,
\lambda_i \gamma_{\mu} \gamma_5 \right) & = & {\cal D} \left( x, x,
\frac{i}{2} \left\{ m, \lambda_i \right\}
\gamma_5 \right), 
\end{eqnarray}
but we also have, as presented elsewhere$^{2)}$,
\begin{eqnarray}
\frac{\partial}{\partial x_{\nu}} {\cal D} \left( x, x, \frac{1}{2} \,
\lambda_i \, \sigma_{\mu \nu} \right) & = & - {\cal D} \left( x, x,
\frac{i}{2} \left\{ m, \lambda_i \right\} \gamma_{\nu} \right)
\nonumber \\   \nonumber \\
& & + \left[ \left( \frac{\partial}{\partial x_{\nu}} -
\frac{\partial}{\partial y_{\nu}} \right) {\cal D} \left( x, y,
\frac{i}{2} \lambda_i \right) \right]_{x=y}
\end{eqnarray}
\\
\begin{eqnarray}
\frac{\partial}{\partial x_{\nu}} {\cal D} \left( x, x, \frac{1}{2} \,
\lambda_i \sigma_{\mu \nu} \gamma_5 \right) & = & - {\cal D} \left(
x, x, \frac{i}{2} \right) \left[ m, \lambda_i \right] 
\gamma_{\nu} \gamma_5
\nonumber \\ \nonumber \\
& & + \left[ \left( \frac{\partial}{\partial x_{\nu}} -
\frac{\partial}{\partial y_{\nu}}  \right) {\cal D} \left(x, y,
\frac{i}{2} \lambda_i \gamma_5 \right) \right] 
\end{eqnarray}
and a number of other formulae, including the following:
\begin{equation}
\left[ \left( \frac{\partial}{\partial x_{\nu}} -
\frac{\partial}{\partial y_{\nu}}  \right) {\cal D} \left( x, y,
\frac{i}{2} \lambda_i \gamma_{\nu} \right) \right]_{x=y} =
{\cal D} \left( x, x, \frac{i}{2} \left\{ \lambda_i, m \right\} \right)
\end{equation}
\\
In the last three formulae, it must be pointed out that for a general
direction of $x-y$ we have the gauge--invariant correspondence
\begin{equation}
{\cal D} \left( x, y, G  \right) \sim \bar{q} (x) Gq (y) \, \, {\rm exp}
\, \, {\rm ig} \, \, \int\limits^{x}_{y} B \cdot dl,
\end{equation}
which is independent of the path from $y$ to $x$ when the coordinate
difference and the path are taken as first order infinitesimals. The first
internal derivative
\begin{equation}
\left[ \left( \frac{\partial}{\partial x_{\mu}} -
\frac{\partial}{\partial y_{\mu}} \right) {\cal D} (x, y, G) \right]_{x=y}
\end{equation}
is physically interesting for all directions $\mu$ (and not just the
-- direction), as a result of Lorentz covariance.\\
In Eqs. (4.5--4.7), we have for the moment thrown off the restriction to
a single null plane.\\
In the next Section, we return to the consideration of the algebra on the
null plane, and we see how further extensions give a much wider algebra,
in which departures from free quark relations begin to appear.\\
\\
\\
\underline{V. THE FURTHER EXTENSION OF NULL PLANE ALGEBRA}\\
\\
We now look beyond the commutation relations of good bilocals on the null
plane. In the model, then, we have to examine operators containing
$q_-$ or $q_-^+$ or both. The Dirac equation in the gauge we are using
($ B_+ = 0$ on the null plane) tells us that we have
\begin{equation}
- 2i \frac{\partial q_-}{\partial x_-} = \left( \alpha_{\perp} \cdot
\left( -i \bigtriangledown _{\perp} - g B_{\perp} \right) + \beta m \right) q_+.
\end{equation}
\\
In terms of Eq. (5.1), we can review the various anticommutators on the
null plane. We have already discussed the trivial one,
\begin{equation}
\left( q_+ (x), q^+_+(y) \right) = \delta \left( x_- - y_- \right).
\frac{1}{2} \left( 1 + \alpha_3 \right) \delta \left( x_{\perp}
- y_{\perp} \right) \, .
\end{equation}
\\
Using (5.1), (5.2), the fact that $B_{\perp}$ commutes with $q_+$ on the
null plane, and the equal--time anticommutator
$\left\{ q_-, q_+^+ \right\} = 0$, we obtain well--known result
\begin{equation}
\left\{ q_-(x), q_+^+ (y) \right\} = \frac{i}{4} \varepsilon
\left( x_- - y_- \right) \left[ \alpha_{\perp}
\cdot \left( i \bigtriangledown_{\perp}^{(y)} - g B_{\perp} (y) \right)
+ \beta m \right] \frac{1}{2} \left( 1 + \alpha_3 \right) \delta
\left( x_{\perp} - y_{\perp} \right) \, .
\end{equation}
Using the same method a second time, one finds, for $y_- > x_-$,
\begin{displaymath}
\left\{ q_-(x), q_-^+ (y) \right\} = - \frac{1}{8} \int^{y_-}_{x_-}
dr_-\left[ \alpha_{\perp} \left( - i \bigtriangledown_{\perp}^{(x)}
- g B_{\perp}  \left( x_{\perp},
r_- \right) \right) + \beta m \right]^2
\left( \frac{1 - \alpha_3}{2} \right) \delta \left( x_{\perp}
- y_{\perp} \right)
\end{displaymath}
\begin{displaymath}
+ i \frac{g^2}{32} \int^{y_-}_{x_-} dy'_- \int^{y'_{-}}_{x_{-}} dx'_-
\left[ \alpha_{\perp} q_+ \left( x_{\perp}, x'_- \right) \right.;
q_+ \left. \left( y_{\perp}, y'_- \right) \alpha_{\perp}
\right] \delta \left( x_{\perp} - y_{\perp} \right)
\end{displaymath}
\begin{equation}
+ \delta \left( x_+ - y_+ \right) \left( \frac{1 - \alpha_3}{2}
\right) \delta \left( x_{\perp} - y_{\perp} \right),
\end{equation}
where the singularity at the coincidence of the two null planes appears
as an unplea\-sant integration constant. This singularity is, of course,
responsible in the model for the Bjorken singularity in the commutator of
two bad or terrible operators.\\
Because of the singularity, it is clumsy to construct the wider algebra
by commuting all our bilocals with one another. Instead, we adopt the
following procedure. Whenever a bilocal operator corresponds to one in the
model containing $q^+_- (x)$, we differentiate with respect to $x_-$;
whenever it corresponds to one in the model containing $q_{(y)}$, we
differentiate with respect to $y_-$. Thus we ``promote'' all our bilocals
to good operators. We construct the wider algebra by starting with the
original good bilocals and these promoted bad and terrible bilocals. We
commute all of these, commute their commutators, and so forth, until the
algebra closes. Then, later on, if we want to commute an unpromoted
operator, we use the information contained in equations of the model like
(5.1) - (5.3) to integrate over $x_- $ or $y_-$ or both and undo the
promotion. (A similar situation obtains for opera\-tors corresponding to
those in the model containing the longitudinal gluon potential $B_-$.)\\
Now let us classify the matrices $\beta G$ into four categories:\\
the good ones, $\beta G = A_{++}$, with $\alpha _3 = 1$ on both sides;\\
the bad ones $\beta G = A_{+-}$ that have $\alpha_3 = 1$ on the left and
$-1$ on the right;\\
the bad ones $\beta G = A_{-+}$ that have $\alpha_3 = -1$ on the left and
$+1$ on the right;\\
and the terrible ones $\beta G = A_{--}$, with $\alpha_3 = -1$ on both
sides.\\
\\
Then, wherever $q_-$ or $q_-^+$ appears, we promote the operator by
differentiating $q_-$ or $q_-^+$ with respect to its argument in the --
direction. We obtain, then:\\
\begin{displaymath}
{\cal D} \left( x, y, \beta A_{++} \right),
\end{displaymath}
the good operators, unchanged;\\
\\
$\frac{\partial}{\partial x_-} \, {\cal D} \left( x, y, \beta A_{-+}
\right)$ and $\frac{\partial}{\partial y_-} \left( x, y, \beta, A_{+-}
\right)$ promoted bad operators:\\
\\
and\\
\\
$\frac{\partial}{\partial x_-} \frac{\partial}{\partial y_-} {\cal D}
\left( x, y, \beta A_{--} \right)$, promoted terrible operators.\\
\\
All 144 of these operators now are given, in the model, in terms of
$q_+$ and $q_+^+$, but the promoted bad and terrible operators involve
the expressions
$\left( \bigtriangledown_{\perp} - ig B_{\perp} \right) q_+$ and
$\left( \bigtriangledown_{\perp} + ig B_{\perp} \right) q_+^+$. In fact, substituting
the Dirac equation for $\frac{\partial q_-}{\partial x_-}$ into the
definitions of the promoted bad and terrible operators, we see that we
obtain good operators (with coefficients depending on bare quark masses)
and also good matrices sandwiched between
$\left( \bigtriangledown_{\perp} + ig B_{\perp} \right) q_+^+$ and $q_+$
or between
$q_+^+$ and $\left( \bigtriangledown_{\perp} -ig B_{\perp} \right) q_+$
or between
$\left( \bigtriangledown_{\perp} + ig B_{\perp} \right) q_+^+$ and
$\left( \bigtriangledown_{\perp} -ig B_{\perp} \right) q_+$.\\
The null plane commutators of all these operators with one another are
finite, well--defined, and physically meaningful, but they lead to an
enormous Lie algebra that is not identical with the one for free quarks,
but instead contains nearly all the degrees of freedom of the model.\\
Let us first ignore any lack of commutation of the B's with one another.
We keep commu\-ting the operators in question with one another. When
$\bigtriangledown_{\perp} \pm ig B_{\perp}$ appears acting on a
$\delta^{(3)}$ function,
we can always perform an integration and fold it over onto an operator.
Thus the number of applications of
$\bigtriangledown_{\perp} \pm ig B_{\perp}$ grows
without limit. Since these gauge derivatives do not commute with one
another, but give field strengths as commutators, it can easily be seen
that we end up with all possible operators corresponding to
$\bar{q}_+ (x) Gq_+ (y)$ acted on by any gauge invariant combination of
transverse gradients and potentials. We have to put it differently, the
operators corresponding to $\bar{q}_+ (x) Gq_+ (y)$ exp ig
$\int\limits_{P} B \cdot dl$ for any pair of points $x$ and
$y$ on the null plane connected by any path $P$ lying in the null plane.
We could think of these as operators ${\cal D} (x, y, G, P)$ depending on
the path $P$, with $\beta G = A_{++}$.\\
\\
In fact the B's do not commute with another in the model, and so we
get an even more complicated result. We have
\begin{equation}
\left[ B_{\perp i} (x), B_{\perp j} (y) \right] \sim \varepsilon
\left( x_- - y_- \right) \delta \left( x_{\perp} - y_{\perp} \right)
\delta_{ij}
\end{equation}
on the null plane, and the commutation of promoted bad and terrible
bilocals with one another leads to operators corresponding to
$\bar{q}_+ (x) Gq_+(y) \bar{q}_+ (a) G'q_+ (b)$. Further commutation
then introduces an unlimited number of sideways gradients, gluon field
strengths, and additional quark pairs, until we end up with all possible
operators of the model that can be constructed from equal numbers of
$\bar{q}_+$'s and $q_+ \, '$s at any points on the null plane and from
exponentials of $ig \int B \cdot dl$ for any paths connecting these points.\\
If we keep track of color, we note that only color singlets are generated.
If the gluons are a color octet Yang--Mills field, we must make suitable
changes in the formalism but again we find that only color singlets are
generated. The coupling constant $g$ that occurs is, of course, the bare
coupling constant. If may not be intrinsic to the algebraic system
(equivalent to that of quarks and gluons) on the null plane, but it
certainly enters importantly into the way we reach the system starting from
well--known operators.\\
A troublesome feature of the extended null plane algebra is the apparent
absence of operators corresponding to those in the model that contain only
gluon field strengths and no quark operators; for a color singlet gluon,
the field strength itself would be such an operator, while for a color
octet gluon we could begin with bilinear forms in the field strength in
order to obtain color singlet operators. Can we obtain these quark--free
operators by investigating discontinuities at the coincidence of
coordinates characterizing quark and antiquark fields in the model? At
any rate, we certainly want these quarkfree operators included in the
extended algebra.\\
Now when our algebra has been extended to include the analogs of all
relevant operators of the model on the null plane that are color singlets
and have baryon number $A = 0$, then the Hilbert space of all physical
hadron states with $A = 0$ is an irreducible representation of the
algebra.\\
If we wish, we might as well extend the algebra further by including the
analogs of color singlet operators of the model (on the null plane) that
would change the number of baryons. In that case, the entire Hilbert
space of all hadron states is an irreducible representation of the
complete algebra. From now on, let us suppose that we are always dealing
with the complete color singlet algebra (whether the one abstracted from
the quark--gluon model or some other) and with the complete Hilbert space,
which is an irreducible representation of it.\\
The representation may be determined by the algebra and the uniqueness
of the physical vacuum. We note that we are dealing with arbitrarily
multilocal operators, functions of any number of points on the null plane.
We can Fourier transform with respect to all these variables and obtain
Fourier variables $\left( k_+, k_{\perp} \right)$ in place of the space
coordinates. Since $B_+ = 0$, there is no formal obstacle to thinking of
each $k_+$ as being like the contribution of the individual quark,
antiquark or gluon to the total $P_+ = \sum k_+$. Now $P_+ = 0$ for the
vacuum, and for any other state we can get $P_+ = 0$ only by taking
$P_z \rightarrow - \infty $. The same kind of smoothness assumption that
allows scaling can allow us to forget about matrix elements to such
infinite momentum states. In that case, we have the unique vacuum state
of hadrons as the only state of $P_+ = 0$, while all others have
$P_+ > 0$. All Fourier components of
multilocal operators for which $\sum k_+ < 0$ annihilate the physical
vacuum. (Note in the null plane formalism we do not have to deal with
a fictitious ``free vacuum'' as in the equal--time formalism.) The Fourier
components of multilocal operator with $\sum k_+ > 0$ act on the vacuum
to create physical states, and the orthogonality properties of these
states and the matrix elements of our operators sandwiched between them
are determined largely or wholly by the algebra. The details have to be
studied further to see to what extent the representation is really
determined. (The vacuum expected values contain one adjustable parameter
in the case of free quarks, namely the number of colors.)\\
Once we have the representation of the complete color singlet algebra
on the null plane, as well as the algebra itself, then the physical states
of hadrons can all be written as linear combinations of the normalized
basis states of the representation. These coefficients represent a
normalized set of Fock space wave functions for each physical hadron state,
with orthogonality relations for orthogonal physical states. Since the
matrix elements of all null plane operators between basis states are
known, the matrix elements between physical states of bilocal currents
or other operators of interest are all calculable in terms of the Fock
space wave functions$^{9)}$.\\
This situation is evidently the one contemplated by ``parton'' theorists
such as Feynman and Bjorken; they suppose that we know the complete
algebra, that it comes out to be a quark--gluon algebra, and that the
representation is the
familiar one, so that there is a simple Fock space of quark, antiquark,
and gluon coordinates. In the Fourier transform, negative values of each
$k_+$ correspond to destruction and positive values to creation.\\
Now the listing of hadron states by quark and gluon momenta is a long
way from listing by meson and baryon moments. However, as long as we
stick to color singlets, there is not necessarily any obstacle to getting
one from the other by taking linear combinations. The operator
$M^2 = - P^2 - P_+ P_-$ has to be such that its eigenvalues correspond to
meson and baryon configurations, and not to a continuum of quarks,
antiquarks and gluons.\\
The important physical questions are whether we have the correct complete
algebra and representation, and what the correct form of $\Theta_{\mu \nu}$
or $P_{\mu}$ or $M^2$ is, expressed in terms of that algebra.\\
In the quark--gluon model we have
$\Theta_{\mu \nu} = \Theta_{\mu \nu}^{{\rm quark}} +
\Theta_{\mu \nu}^{{\rm glue}}$, where
\begin{eqnarray}
\Theta^{{\rm quark}}_{\mu \nu} & = & \frac{1}{4} \bar{q} \gamma_{\mu}
\left( \partial_{\nu} - ig B_{\nu} \right) q + \ldots
q + \frac{1}{4} \bar{q} \gamma_{\nu} \left( \partial_{\mu} - ig
B_{\mu} \right) q \nonumber \\ \nonumber \\
& & - \frac{1}{4} \left( \partial_{\mu} + ig B_{\mu} \right) \bar{q}
\gamma_{\nu} q - \frac{1}{4} \left( \partial_{\nu} + ig B_{\nu} \right)
\bar{q} \gamma_{\mu} q \, ,
\end{eqnarray}
and $\Theta_{\mu \nu}^{{\rm glue}}$ does not involve the quark variables at
all. The term $\Theta_{\mu \nu}^{{\rm quark}}$, by itself, has the wrong
commutation rules to be a true $\Theta_{\mu \nu}$ (unless $g = 0$). For
example, $\left( P_1^{{\rm quark}}, P_2^{{\rm quark}} \right) \not= 0$.
The correct commutation rules are restored when we add the contribution
from $\Theta_{\mu \nu}^{{\rm glue}}$.\\
We can abstract from the quark--gluon model some or all the properties
of $\Theta_{\mu \nu}$, in terms of the null plane algebra. We see that
in the model we have
\begin{equation}
\Theta_{++}^{{\rm quark}} = \left[ \left( \frac{\partial}{\partial y_-}
- \frac{\partial}{\partial x_-} \right) {\cal D} \left( x, y,
\frac{1}{2} \gamma_+ \right) \right]_{x=y}
\end{equation}
\\
and, as is well--known, the expected value of the right--hand side in
the proton state can be measured by deep inelastic experiments with
electrons and neutrinos. All indications are that it is not equal to the
expected value of $\Theta_{++}$, but rather around half of that, so that
half is attributable to gluons, or whatever replaces them in the real
theory.\\
\\
In general, using the gauge--invariant definition of ${\cal D}$, we have
in the model
\begin{equation}
\Theta_{\mu \nu}^{{\rm quark}} = \left[ \left(
\frac{\partial}{\partial y_{\nu}} -  \frac{\partial}{\partial x_{\nu}}
\right) {\cal D} \left( x, y, \frac{1}{4} \gamma_{\mu} \right) \right.
 + \left. \left( \frac{\partial}{\partial y_{\mu}} -
\frac{\partial}{\partial x_{\mu}} \right) {\cal D} \left( x, y,
\frac{1}{4} \gamma_{\nu} \right) \right]_{x=y}
\end{equation}
and Eq. (4.7) then gives us the obvious result
\begin{equation}
- \Theta^{{\rm quark}}_{\mu \nu} = {\cal D} \left( x, x, m \right) \, .
\end{equation}
\\
Whereas in (5.5) we are dealing with an operator that belongs to the null
plane algebra generated by good, promoted bad, and promoted terrible
bilocal currents, other components of $\Theta_{\mu \nu}^{{\rm quark}}$
are not directly contained in the algebra, neither are the bad and
terrible local currents, nor their internal derivatives in directions
other than --. In order to obtain the commutation properties of all these
operators with those actually in the algebra, we must, as we mentioned
above, undo the promotions by abstracting the sort of information
contained in (5.3) and (5.4). Thus we are really dealing with a wider
mathematical system than the closed Lie algebra abstracted from that of
operators in the model containing $q^+_+, q_+$ and $B_{\perp}$ only.\\
\\
We shall assume that the true algebraic system of hadrons resembles that
of the quark--gluon model at least to the following extent:
\begin{enumerate}
\item[1)] The null plane algebra of good components (4.2) and the leading
light cone singularities (4.1) are unchanged.

\item[2)] The system acts as if the quarks had vectorial coupling in
the sense that the divergence equation (4.3) and (4.4) are unchanged.

\item[3)] There is a gauge derivative of some kind, with path--dependent
bilocals that for an infinitesimal interval become path--independent.
Eqs. (4.5) - (4.7) are then defined and we assume they also are unchanged.

\item[4)] The expression (5.6) for $\Theta_{\mu \nu}^{{\rm quark}}$ is
also defined and we assume it, too, is unchanged, along with its corollary
(5.7).
\end{enumerate}
About the details of the form of the path--dependent null plane algebra
arising from the successive application of gauge derivatives, we are much
less confident, and correspondingly we are also less confident of the
nature of the gluons, even assuming that we can decide whether to use a
color singlet or a color octet. What we do assert is that there is some
algebraic structure analogous to that in quark--gluon theory and that it
is in principle knowable.\\
One fascinating problem, of course, is to understand the conditions under
which we can have an algebra resembling that for quarks and gluons and yet
escape having real quarks and gluons. Under what conditions do the
bilocals act as if they were the products of local operators without, in
fact, being seen. We seek answers to this and other questions by asking
``Are there models other than the quark--gluon field theory from which
we can abstract results? Can we replace $\Theta_{\mu \nu}^{{\rm glue}}$
by something different and the gauge--derivative by a different
gauge--derivative?''\\
\\
\\
\underline{VI. ARE THERE ALTERNATIVE MODELS?}\\
\\
In the search for alternatives to gluons, one case worth investigating is
that of the simple dual resonance model. It can be considered in three
stages: first, the theory of a huge infinity of free mesons of all spins;
next, tree diagrams involving the interaction of these mesons; and finally
loop diagrams. The theory is always treated as though referring to real
mesons, and an $S$--matrix formulation is employed in which each meson
is always on the mass shell.\\
Now the free stage of the model can easily be reformulated as a field
theory in ordinary coordinate space, based on a field operator $\Phi$
that is a function not of one point in space, but of a whole path -- it
is infinitely multilocal. The free approximation to the dual resonance
model is then essentially the quantum theory of a relativistic string or
linear rubber band in ordinary space.\\
The coupling that leads, on the mass shell, to the tree diagrams of the
dual resonance model has not so far been successfully reformulated as a
field theory coupling but we shall assume that this can be done. Then
the whole model theory, including the loops, would be a theory of a large
infinity of local meson fields, all described simultaneously by a grand
infinitely multilocal field $\Phi$, couples to themselves and one another.
The mesons, in the free approximation, lie on straight parallel Regge
trajectories with a universal slope $\alpha '$.\\
In the simplest form of such a theory, the grand field $\Phi$ (path) can
be resolved into local fields $\phi (R), \Phi_{n \mu} (R),
\Phi_{n \mu, n' \mu'} (R), \ldots $. There is a single scalar, a single
infinity of vectors, a double infinity of tensors and scalars, and so
forth. The matrices $a_{n \mu}$ and $a_{n \mu}^+$ of the dual theory
connect these components of $\Phi$ with one another.\\
Perhaps the model theory of a gluon field can be replaced by a field
theory version of a dual resonance model; the properties of operators,
including  $\Theta_{\mu \nu}$, would be abstracted from the new model
instead of the old one. With $\alpha ' \not= 0$, a term $\delta $ would
naturally appear that violates scale invariance and is not related to the
bare quark masses. (Probably $l_{\delta } = 0$ here rather than $-2$ as
in the case of a gluon mass.) The gauge derivative in the other portion
of $\Theta_{\mu \nu} $, referring to the quarks, would then involve a
special linear combination of the $\Phi_{n \nu} (R)$ instead of the gluon
potential $B_{\mu} (R)$.\\
An amusing point is that in the limit of a dual resonance theory as
$\alpha ' \rightarrow 0$ (so that the trajectories become flat), with
attention concentrated on the value $\alpha = 1$, if the mathe\-matics of
a Lie group is built into the model, then the mass shell predictions become
those of the corresponding massless Yang--Mills theory$^{10)}$. That suggests
that one might even try a dual resonance model as a replacement of a color
octet Yang--Mills gluon model, with abstraction of the properties of color
singlet operators.\\
We are not at all sure that what we are discussing here is a practical
scheme, and if it is, we do not know how the resulting algebraic system
differs from that of gluons. We put it forward merely in order to
stimulate thinking about whether or not here are candidates for the
algebra, the representation, and the form of $\Theta_{\mu \nu}$ other
than those suggested by the gluon model.\\
Our attempt to use the dual model to construct a field theory has no
bearing on whether the mass--shell dual model can lead to a complete
$S$--matrix theory of hadrons; our suggestion resembles the use of limits
of dual theories to obtain unified theories of weak and electromagnetic
interactions or the theory of gravity.\\
One interesting speculation that is independent of what model we use for
the stuff to which quarks are coupled is that perhaps when we perform
the mathematical transformation from current quarks to constituent
quarks and obtain the crude naive quark model of meson and baryon spectra
and couplings, the gluons or whatever they are will also be approximately
transformed into fictitious constituents, so that meson states would
appear that act as if they were made of gluons rather than $q \bar{q}$
pairs. If there are indeed ten low--lying scalar mesons rather than nine,
then we might interpret the tenth one (roughly speaking, the
$\varepsilon ^{\circ}$ meson) as the beginning of such a sequence of
extra $Su_3$ singlet meson states. (A related question, much debated by
specialists in the usual, mass--shell dual models, is whether the infinite
sequence of meson and baryon Regge trajectories, all rising indefinitely
and straight and parallel in zeroth approximation, should be extended
to exotic channels, i. e., those with quantum numbers characteristic
of $qqqq \bar{q}, q \bar{q} q \bar{q}$ etc.).\\
Let us end by emphasizing our main point, that it may well be possible
to construct an explicit theory of hadrons, based on quarks and some kind
of glue, treated as fictitious, but with enough physical properties
abstracted and applied to real hadrons to constitute a complete theory.
Since the entities we start with are fictitious, there is no need for
any conflict with the bootstrap or conventional dual model point of view.\\
\newpage
\noindent
\begin{center}
APPENDIX -- BILOCAL FORM FACTOR ALGEBRA
\end{center}
\bigskip
We have described in Section III and IV a Lie algebra of good components
of bilocal operators on a null plane. The generators are 36 functions of
$x_-, y_-$ and $x_{\perp} = y_{\perp}$, namely ${\cal F}_{j+},
{\cal F}_{j+}^5, {\cal T}_{jl+}$, and ${\cal T}_{j2+}$. We define
$R \equiv 1/2 (x + y)$ and $z \equiv x-y$; then we have functions of
$R_{\perp}, R_-$, and $z_-$.\\
With $z_-$ set equal to zero, we have just the usual good local operators
on the null plane, related to the corresponding good local operators at
equal times with $P_3 \rightarrow \infty$. We recall that in the early
work using $P_3 \rightarrow \infty$ the most useful applications (fixed
virtual mass sum rules) involved matrix elements with no change of
longitudinal momentum, i. e., transverse Fourier components of the
operators. Dashen and Gell--Mann$^{11)}$ studied these operators and found
that between finite mass states their matrix elements do not depend
separately on the transverse momenta of the initial and final states, but
only on the difference, which is the Fourier variable $k_{\perp}$. Thus
they obtained a ``form factor algebra'' generated by operators
$F_i \left( k_{\perp} \right)$ and $F_i^5 \left( k_{\perp} \right)$, to
which, of course, one may adjoin $T_{il} \left( k_{\perp} \right)$ and
$T_{i2} \left( k_{\perp} \right)$.\\
We may consider the analogous quantities using the null plane method and
generating to bilocals:\\
\\
$F_i \left( k_{\perp}, z_- \right)  \equiv$\\
\begin{equation}
\int d^4 R \delta \left(
R_+ \right) {\cal F}_{i+} \left( R,
z_- \right) \, \, \, {\rm exp} \, \, \, i k_1 \left[ R_1 + P_+^{-1}
\left( \Lambda_1 + J_2 \right) \right]
\, \, \, {\rm exp} \, \, \, i k_2 \left[ R_2 + P_+^{-1} \left(
\Lambda_2 - J_1 \right) \right]
\end{equation}
and so forth. Here the integration over $R_-$ assures us that
$P_+ \equiv P_0 + P_3 $ is conserved by the operator. (We note that
Minkowski$^{12)}$ and others have studied the interesting problem of
extracting useful sum rules from operators unintegrated over
$R_-$, but we do not discuss that here.) The quantities
$P_+^{-1} \left( \Lambda_1 + J_2 \right)$ and $P_+^{-1} \left( \Lambda_2
- J_1 \right)$ act like negatives of center--of--mass coordinates,
$- \bar{R}_1$ and $-\bar{R}_2$, since on the null plane $x_+ = 0$ we have
$\Lambda_1 + J_2 = - \int R_1 \Theta_{++} d^4 R \delta \left( R_+ \right)$ and
$\Lambda_1 + J_1 = - \int R_2 \Theta_{++} d^4 R \delta \left( R_+ \right)$,
while $P_+ = \int \Theta_{++} d^4 R \left( R_+ \right)$.\\
Our bilocal form factor algebra has the commutation rules
\begin{equation}
\left[ F_i \left( k_{\perp}, z_- \right), F_j \left( k_{\perp}', z_-'
\right) \right] = i \, f_{ijk} F_k \left( k_{\perp} + k_{\perp}', z_-
+ z_-' \right),
\end{equation}
etc., where the structure constants in general are those of $\left[ U_6
\right]_w$. Putting $z_- = z_- \, ' = 0$, we obtain exactly the form factor
algebra of Dashen and Gell--Mann. If we specialize further to
$k_{\perp} = k_{\perp}' = 0$, we obtain the algebra
$\left[ U_6 \right]_{w, \infty, \, \, \, {\rm currents}}$, of vector,
axial vector, and tensor charges. It is not, of course, identical to the
approximate symmetry algebra
$\left[ U_6 \right]_{w, \infty \, \, \, {\rm strong}}$, for baryon and
meson spectra and vertices, but is related to it by a transformation,
probably unitary. That is the transformation which we have described
crudely as connecting current quarks and constituent quarks.\\
The behavior of the operators $F_i \left( k_{\perp} \right)$, etc., with
respect to angular momentum in the s--channel is complicated and
spectrum--dependent; it was described by Dashen and Gell--Mann in their
angular condition$^{10)}$. There is a similar angular condition for the
bilocal generalizations $F_i \left( k_{\perp}, z_- \right)$, etc.\\
The behavior of $F_i \left( k_{\perp}, z_- \right)$ and the other bilocals
with respect to angular momentum in the cross--channel is, in contrast,
extremely simple. If we expand $F_i \left( k_{\perp}, z_- \right)$ or
$F_i^5 \left( k_{\perp}, z_- \right)$ in powers of $z_-$, each power
$z_-^n$ corresponds to a single angular momentum, namely $J = n + 1$.\\
As we expand $F_i \left( k_{\perp}, z_- \right)$, etc., in power series in
$z_-$, we note that each term, in $z_- \, ^{J-1}$, has a pole in
$k_{\perp}^2$ at $k_{\perp}^2 + M^2 = 0$, where $M$ is the mass of any
meson of spin $J$. By an extension of the Regge procedure, we can keep
$k_{\perp}^2$ fixed and let the angular momentum vary by looking at the
asymptotic behavior of matrix elements of $F_i \left( k_{\perp}, z_-
\right)$, etc., at large $z_-$. A Regge pole in the cross channel gives
a contribution
$z_- ^{\alpha \left( - k_{\perp}^2 \right)} \beta \left( k_{\perp}^2
\right) \, \left[{\rm sin} \, \, \, \pi \alpha \left(-k_{\perp}^2 \right)
\right]^{-1}$
and a cut gives a corresponding integral over $\alpha $.
Thus the bilocal form factor $F_i \left( k_{\perp}, z_- \right)$ couples
to each Reggeon in the non--exotic meson system in the same way that
${\cal F}_i \left( k_{\perp} \right)$ couples to each vector meson. The
contribution of each Regge pole to the asymptotic matrix element of
$F_i \left( k_{\perp}, z_- \right)$ between hadron states $A$ and $B$ is
given by the coupling of ${\cal F}_i \left( k_{\perp}, z_- \right)$ to
that Reggeon multiplied by the strong coupling constant of the
Reggeon to $A$ and $B$.\\
It would be nice to substitute the Regge asymptotic behavior of
$F_i \left( k_{\perp}, x_- \right)$ etc., into the commutation rules and
obtain algebraic relations among the Regge residues. Unfortunately, the
asymptotic limit is not approached uniformly in the different matrix
elements, and the asymptotic Regge formulae cannot, therefore, be used
for the operators
everywhere in the equations (A.2); only partial results can be extracted.
\newpage
\noindent
References
\begin{enumerate}
\item[1.] M. Gell--Mann, Phys. Rev. \underline{125}, 1067 (1962) and
Physics \underline{1}, 63 (1964).

\item[2.] H. Fritzsch, M. Gell--Mann, Proceedings of the Coral Gables
Conference on Fundamental Interactions at High Energies, January 1971,
in ``Scale Invariance and the Light Cone'', Gordon and Breach Ed. (1971),
and Proceedings of the International Conference on Duality and Symmetry
in Hadron Physics, Weizmann Science Press (1971).\\
J. M. Cornwall, R. Jackiw, Phys. Rev. \underline{D4}, 367, (1971).\\
C.H. Llewellyn Smith, Phys. Ref. \underline{D4}, 2392, (1971).

\item[3.] D. J. Gross, S. B. Treiman, Phys. Rev. \underline{D4}, 1059,
(1971).

\item[4.] M. Gell--Mann, Schladming Lectures 1972, CERN--preprint TH 1543.\\
W. A. Bardeen, H. Fritzsch, M. Gell--Mann, Proceedings of the Topical
Meeting on Conformal Invariance in Hadron Physics, Frascati, May 1972.

\item[5.] J. Wess (Private communication to B. Zumino).

\item[6.] H. Fritzsch, M. Gell--Mann and A. Schwimmer, to be published.\\
D. J. Broadhurst and R. Jaffe, to be published.

\item[7.] H. Leutwyler, J. Stern, Nuclear Physics \underline{B20}, 77
(1970).

\item[8.] R. Jackiw, DESY Summer School Lectures 1971, preprint MIT--CTP
236.

\item[9.] G. Domokos, S. K\"ovesi--Domokos, John Hopkins University
preprint C00--3285--22, 1972.

\item[10.] A. Neveu, J. Scherk, Nuclear Physics \underline{B36}, 155, 1972.

\item[11.] R. Dashen, M. Gell--Mann, Phys. Rev. Letters \underline{17},
340 (1966).\\
M. Gell--Mann, Erice Lecture 1967, in: Hadrons and their Interactions,
Academic Press, New York--London, 1968.\\
S.--J. Chang, R. Dashen, L. O'Raifeartaigh, Phys. Rev. \underline{182},
1805 (1969).

\item[12.] P. Minkowski, unpublished (private communication).
\end{enumerate}
\end{document}